\documentclass[10pt,a4paper]{article}
\usepackage{multicol}
\usepackage{amssymb,amsmath,latexsym}
\usepackage[USenglish]{babel}
\usepackage[runin]{abstract}
\newcounter{myFCounter}[section]
\usepackage[pdftex]{graphicx} 
\usepackage[labelfont = bf, labelsep = period, font = it, aboveskip = 0pt, belowskip = 0pt]{caption}
\makeatletter
\renewcommand{\thesection}{\@arabic\c@section}
\renewcommand{\thesubsection}{\thesection.\@arabic\c@subsection}
\renewcommand{\section}{\@startsection {section}{1}{\parindent}%
    {3.5ex \@plus 1ex \@minus .2ex}%
    {2.3ex \@plus .2ex}%
    {\normalfont\Large\bfseries}}
\renewcommand{\subsection}{\@startsection {subsection}{2}{\parindent}%
    {3.25ex \@plus 1ex \@minus .2ex}%
    {1.5ex \@plus .2ex}%
    {\normalfont\large\bfseries}}
\renewcommand{\paragraph}{\@startsection {paragraph}{4}{\parindent}%
    {3.25ex \@plus1ex \@minus.2ex}%
    {-1em}%
    {\normalfont\normalsize\bfseries}}
\renewcommand{\tableofcontents}{%
    \section*{\hspace{-\parindent}\contentsname
    \@mkboth{\MakeUppercase\contentsname}{\MakeUppercase\contentsname}}%
    \@starttoc{toc}}

\renewenvironment{thebibliography}[1]
     {\section*{\hspace{-\parindent}\refname}%
      \@mkboth{\MakeUppercase\refname}{\MakeUppercase\refname}%
      \list{\@biblabel{\@arabic\c@enumiv}}%
           {\settowidth\labelwidth{\@biblabel{#1}}%
            \leftmargin\labelwidth
            \advance\leftmargin\labelsep
            \@openbib@code
            \usecounter{enumiv}%
            \let\p@enumiv\@empty
            \renewcommand\theenumiv{\@arabic\c@enumiv}}%
      \sloppy
      \clubpenalty4000
      \@clubpenalty \clubpenalty
      \widowpenalty4000%
      \sfcode`\.\@m}
     {\def\@noitemerr
       {\@latex@warning{Empty `thebibliography' environment}}%
      \endlist}

\newcommand{\vv}{\vec{v}}

\newcommand{\vr}{\vec{r}}

\newcommand{\vE}{\vec{E}}

\newcommand{\vB}{\vec{B}}

\newcommand{\partDer}[2]{\frac{\partial #1}{\partial #2}}

\newcommand{\dfull} [2]{\frac{d #1}{d #2}}

\begin {document}
\title{\textbf{\textsc{Single-Species Weibel Instability \\of Radiationless Plasma}}}
\author{L.\,V. Borodachev$^{a}$, D.\,O. Kolomiets$^{b}$ \\
\\
\small{Department of Mathematics, Faculty of Physics,}\\
\small{M.V. Lomonosov Moscow State University, Moscow 119991, Russia.}\\
\small{E-mail: $^a$borodach2000@mail.ru, $^b$kolomiets@darwincode.org}}
\setlength{\abstitleskip}{-\absparindent} \abslabeldelim{.\;}
\date{\small{June 2010}\\
} \maketitle \hrule
\begin{abstract}
A Particle-in-Cell (PIC) numerical simulation of the electron Weibel instability is applied in a
frame of Darwin (radiationless) approximation of the self-consistent fields of sparse plasma. As a
result, we were able to supplement the classical picture of the instability and, in particular, to
obtain the dependency of the basic characteristics (the time of development and the maximum field
energy) of the thermal anisotropy parameter, to trace the dynamic restructuring of current
filaments accompanying the nonlinear stage of the instability and to trace in detail the evolution
of the initial anisotropy of the electron component of plasma.\\
\\
\textit{Keywords:} Weibel instability, Particle-in-Cell, PIC, Vlasov--Darwin model. \\
\textit{PACS:} 52.35.Qz.
\end{abstract}
\hrule \setlength{\abstitleskip}{-\absparindent} \abslabeldelim{.\;} \maketitle

\newcommand{\intbt}[2]{\int\limits_{#1}^{\mspace{6mu} #2}\mspace{-5mu}}
\begin{multicols}{2}
\section*{Introduction}\label{S:Intro}

As it is well known, anisotropic distribution of thermal velocities in homogeneous collisionless
plasma may lead to Weibel instability (WI) in case of spontaneous generation of transverse to
direction of anisotropy magnetic field \cite{Weibel.article}. The peculiarity of WI is to have a
regime where unstable growth of magnetic field generated by arisen current layers (filaments in
multidimensional case) is followed by dynamic transformation of the current structures and
stabilization of the magnetic field energy after saturation of the instability.

The general picture of Weibel instability mentioned above has significant variations in several of
practically important applications of plasma physics, e.g. WI in relativistic plasma which absorbed
high-energy EM impulse \cite{Pukhov} is considerably different comparing to WI occurring in the
current sheet of Earth's magnetotail \cite{Yoon.ionWI}, which in turn differs from WI of intense
ion beams \cite{Davidson.MWI}. Recent researches indicate that WI might have an important role in
dense quantum plasmas \cite{Tsintsadze.Shukla}, where WI is supposed to be responsible for the
generation of non-stationary magnetic fields in compact astrophysical objects as well as in laser
fusion experiments.

That variety of forms along with fundamental nature of WI stimulate its further study using
different approaches, one of which is a PIC-method based on Vlasov--Darwin model
\cite{Borodachev.model} applied below. Presence of well-developed linear theory of electromagnetic
instabilities \cite{Mikhaylovskiy.plasma}, a typical representative of which is WI, allows one to
verify the obtained numerical results.

This work is devoted to the numerical simulation of the electron Weibel instability in order to
clarify its general picture, in particular, the evolution of the initial anisotropy on the time,
and the dependency of key instability characteristics of the initial anisotropy. Ions due to
inertia are considered to be a neutralizing positive background.

\section{Elements of linear theory \\ of Weibel instability}\label{S:WeibelLinear}
In order to get notion of the generation mechanism of the Weibel instability we will review the
basic results of linear theory applied to it in the most simple, 1D3V ($x, v_x, v_y, v_z$), setup.

Let's consider a homogeneous collisionless plasma with a positive (ion) background and anisotropic
distribution of thermal velocities of electrons, given by the following distribution function:
\begin{equation}
    f_0\left(\vec{v}\right) = \frac{n_0\exp\left(-\frac{v_x^2}{u_x^2} -
    \frac{v_y^2}{u_y^2} - \frac{v_z^2}{u_z^2}\right)}{\pi^{3/2}u_x u_y u_z},
    \label{E:weibelF0}
\end{equation}
where $n_0$ is the density of electrons, $u_x, u_y, u_z$ are the thermal velocities of the
electrons along the corresponding axes, where we assume (without loss of generality) $u_z > u_x =
u_y$.

In this case the plasma, obviously, has an excess of fast particles moving along $z$-axis. However,
due to the symmetry of the distribution, the total current is zero.

Now assume that there exists spontaneously generated (due to thermal fluctuations) non-zero
magnetic field $\vec{B}_0 = \vec{e}_y B_{y0} \sin\left(k_x x\right)$. Then the Lorentz force
$\vec{F}_L = -e\vec{v}\times\vec{B}_0/c$ will change the trajectory of a particle moving along $z$,
which will lead to formation of spatially separated current sheets. Localization of these sheets
will coincide with the zeros of the magnetic field which caused the deflection of the particles.
Hence, we should expect formation of two current sheets (with opposite signs) per each wavelength
of the magnetic field perturbation, amplifying the initial magnetic field. The growth of the field,
which in turn increases density of the current sheets, will continue until the major part of the
particles become significantly magnetized. Fig.\:\ref{F:Explanation3D} shows a qualitative
illustration of this process.
\begin{center}
    \begin{minipage}[t]{\columnwidth}
        \refstepcounter{figure}
        \begin{center}
            \includegraphics[width=\columnwidth,keepaspectratio]
            {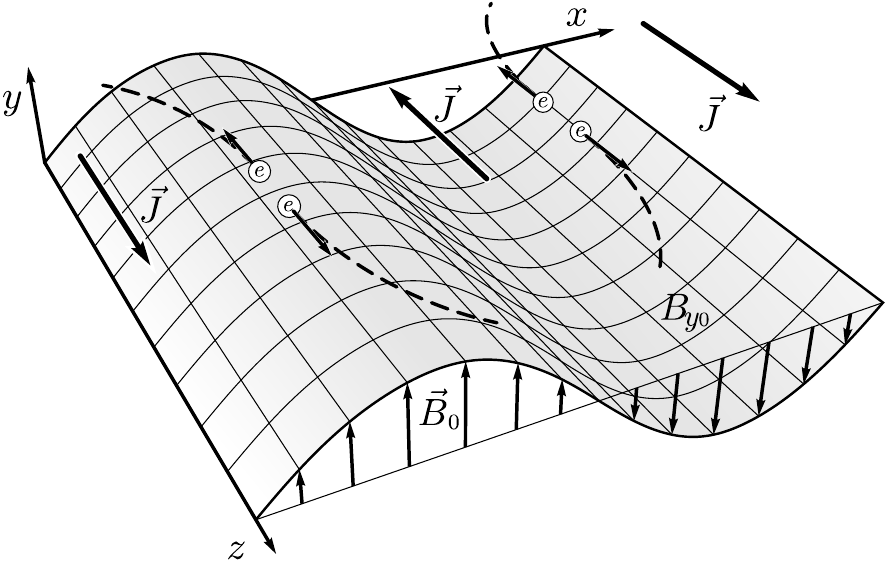}\label{F:Explanation3D}
        \end{center}
        \textit{Fig.\:\arabic{figure}.
        Qualitative representation of the linear phase of Weibel instability. Development of
        current sheets $\vec{J}$ around the zeros of $B_y$ (shown as a surface). $B_x$, $B_z$
        considered zero for simplicity.}
    \end{minipage}
\end{center}

In the described setup the perturbed quantities are $E_z(x,t)$, $B_y(x,t)$ (unperturbed values of
which are zero) and the electron distribution function $f(x,\vec{v},t)  = f_0(\vec{v}) + f_1(x,
\vec{v}, t)$.

Let the perturbations (as it's common for linear analysis in general) to be represented as
$\exp\left(i\left(k_x x - \omega t\right)\right)$. Then, substituting the corresponding values into
the Vlasov equation and the field equations, we obtain (as it is shown in \cite{Morse.Nielson}) a
dispersion equation
\begin{equation}\label{E:Dispersion}
    k_x^2c^2 - \omega^2 = \omega_{\mathrm{p}e}^2 \left(A + \frac{\omega\left(A + 1 \right)}{k_x
    u_x}\:Z\!\left(\frac{\omega}{k_x u_x}\right)\right),
\end{equation}
where $\omega_{\mathrm{p}e} = \sqrt{4\pi n_0 e^2/m}$ is the electron plasma frequency, $A =
(u_z^2/u_x^2 - 1)$ is the anisotropy parameter and $Z$ is the plasma dispersion function
\cite{Frank}:
\begin{equation*}
    Z\left(\zeta\right) = \frac{1}{\sqrt{\pi}}\intbt{-\infty}{\infty}\frac{e^{-s^2}}{s-\zeta}ds,
\end{equation*}
where $\zeta$ in general is a complex value.

The unstable roots of Eq.\:(\ref{E:Dispersion}) (solutions with purely imaginary positive $\omega$,
the standing waves) are in the range from $k c/\omega_{\mathrm{p}e} = 0$ to $k
c/\omega_{\mathrm{p}e} = \sqrt{A}$ for any $A>0$. Thus in the case when the dimensionless wave
number $k_x$ falls into this interval, there is a sharp increase in the amplitude of the
perturbation of the corresponding wavelength, which is a key feature of the initial stage of WI.
The growth rate $\gamma = i\omega$ has a single maximum in the above-mentioned range, which allows
to determine the most unstable mode. Note that it's hard (if at all possible) to get analytical
solution of this equation, but it is rather easy to solve it numerically with any standard method
\cite{Kalitkin}. The foregoing is illustrated in Fig.\:\ref{F:GammaMultiMode} where the graphs of
the growth rate are shown for various values of anisotropy parameter with fixed accentuated ($u_z$)
thermal velocity.
\begin{center}
    \begin{minipage}[t]{\columnwidth}
        \refstepcounter{figure}
        \begin{center}
            \includegraphics[width=\columnwidth,keepaspectratio]
            {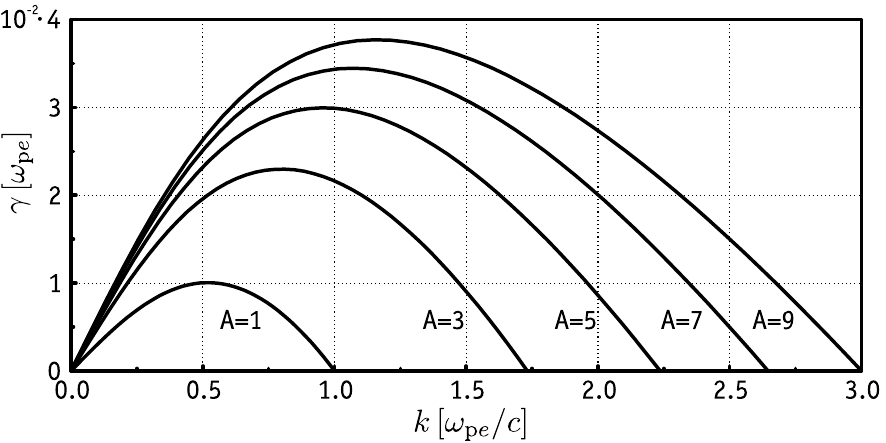}\label{F:GammaMultiMode}
        \end{center}
        \textit{Fig.\:\arabic{figure}.
        Growth rate of WI for various values of anisotropy at constant accentuated thermal velocity
        $u_z=0.1\,[c].$}
    \end{minipage}
\end{center}

\section{Numerical simulations}
Here we will briefly review the basic features of the model we chose, as well as it's geometrical
setup and the main parameters.

The general physical properties of the system were already described above in the section
\ref{S:WeibelLinear}. Now we will consider 2.5-dimensional setup ($x,y,v_x,v_y,v_z$), where the
accentuated component of the thermal velocity is taken perpendicularly to the simulation plane
$x$-$y$. Note that in this case the development of the instability will now be supported with
formation of current filaments (beams), rather than layers, as it was in 1.5-dimensions. However,
the results of the linear theory presented above are still valid in this case, if one assumes $k_x
= k_y = k$.

First of all, we should note that low-frequency nature of the WI itself allows us to use Darwin
(radiationless) approximation of the electromagnetic fields \cite{Darwin}, which is significantly
more efficient from computational point of view comparing to more common full Maxwell model.

Thus, in the PIC-simulations performed here, the evolution of plasma system inside of the
computational domain described by the dynamic equations of macroparticles (electrons)
\begin{equation}\label{E:fieldDynamics}
    \left\{
        \begin{aligned}
            &\dfull{\vv_p}{t} = \frac{q_p}{m_p} \left(\vE(\vr_p, t) +
            \frac{\vv_p \times \vB(\vr_p, t)}{c}\right),\\
            &\dfull{\vr_p}{t} = \vv_p, \quad p = 1, 2, \ldots ,N_{\mathrm{p}},
        \end{aligned}
    \right.
\end{equation}
moving in the self-consistent fields determined by the Darwin's equations
\begin{equation}\label{E:fieldDynamics}
    \left\{
        \begin{aligned}
            &\nabla\times\vec{B} = \frac{4\pi}{c}\vec{J} +
            \frac{1}{c}\partDer{\vec{E}_l}{t},\\
            &\nabla\vec{B} = 0,\\
            &\nabla\times\vec{E}_t =
            -\frac{1}{c}\partDer{\vec{B}}{t},\\
            &\nabla\vec{E}_l = 4\pi\rho,\\
            &\vec{E} = \vec{E}_l + \vec{E}_t, \quad \nabla\times\vec{E}_l = 0, \quad \nabla\vec{E}_t = 0,
        \end{aligned}
    \right.
\end{equation}
where $\vec{E}_l$ and $\vec{E}_t$ are, correspondingly, the longitudinal (curl-free) and transverse
(divergence-free) components of the electric field.

As it is easy to see, Darwin's approximation differs from the full Maxwell description only by the
eliminated transverse displacement current, which physically means neglecting radiation and
transition to fields with instantaneous propagation \cite{Nielson}. At the same time the system
partially keeps inductive effects (associated with Faraday's law) and ensures holding of the
continuity equation due to the presence of longitudinal part of the displacement current.

Then we specify the uniform spatial distribution of electrons and singly charged ions, the latter
ones are considered as a motionless neutralizing background. In the expression (\ref{E:weibelF0})
we choose $u_z = 0.1\,[c]$, $u_x = u_y = 0.0316\,[c]$. For these values of thermal velocities the
anisotropy parameter equals to $A_0 = 9$.

The characteristic linear size of the computational domain, which we can choose for the simulation,
must correspond to the wavelength for which the dispersion equation (for given values of anisotropy
and accentuated thermal velocity) shows the instability growth rate close to maximum
(Fig.\:\ref{F:GammaMultiMode}).

In particular, solving numerically the equation (\ref{E:Dispersion}) for our reference case $A_0 =
9$ and $u_z = 0.1\,[c]$, we obtain the approximate value of $\gamma_\mathrm{max} = 0.037$ and the
corresponding value of $k_\mathrm{max} = 1.2 \, [\omega_{\mathrm{p}e}/c]$, in terms of the
wavelength it will give us $\lambda_\mathrm{max} = 5.2 \, [c/\omega_{\mathrm{p}e}]$.

Thus, choosing a rectangular computational domain $L_x = L_y = 25 \, [c/\omega_{\mathrm{p}e}]$ will
allow us to trace in detail the development of initial current system and its restructuring in
nonlinear stage of WI.

Determination of the total number of particles $N_{\mathrm{p}}$ is preconditioned by a natural
trade-off between holding colisionless property of the model in the time-frame sufficient for
development of the instability and computational cost of a singe run. Using theoretical estimations
of the collisionless time for big particles (with width $W > \lambda_D$) \cite{Hockney.book}, as
well as test runs, it was found that the number of particles $N_{\mathrm{p}}$ of order $10^6$
allows to have the collisionless period at least 5 times longer than it is needed for the
development of WI.

Following the work \cite{Morse.Nielson}, we choose periodic boundary conditions for both $x$ and
$y$ directions.

Finally, note that in the series of experiments considered below variation of the the initial
anisotropy parameter $A_0$ has been done by changing of $u_x$ and $u_y$ keeping $u_z$ constant in
all runs to keep non-relativistic regime of the simulations.

\section{Discussion of results}
The instability development process is clearly visible on the plots of the average over space
magnetic field energy density versus time for different initial values of the anisotropy shown in
Fig.\:\ref{F:WebelStepsLnBEnergy}.
\begin{center}
    \begin{minipage}[b]{\columnwidth}
        \refstepcounter{figure}
        \begin{center}
            \includegraphics[width=\columnwidth,keepaspectratio]
            {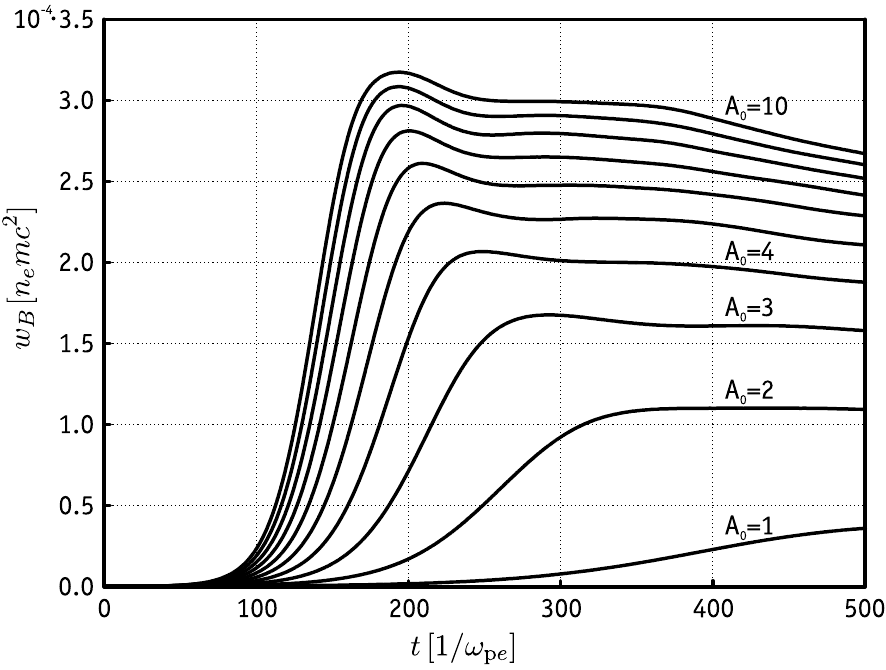}\label{F:WebelStepsLnBEnergy}
        \end{center}
        \textit{Fig.\:\arabic{figure}.
        Average magnetic field energy density versus time for different values of initial anisotropy ($A_0 = 1...10$)
        at constant accentuated thermal velocity. Computational domain $L_x = L_y =
        25\,[c/\omega_{\mathrm{p}e}]$; $u_z = 0.1\,[c]$; mesh size $128\times128$; 500 particles of each specie per cell;
        time step $\tau = 0.25\,[1/\omega_{\mathrm{p}e}].$}
    \end{minipage}
\end{center}
The initial value of the average energy density is close to zero, but by the time $t =
100\,[1/\omega_{\mathrm{p}e}]$, corresponding to appearing of distinguishable areas of current
localization, it is observed a noticeable increase, which is the greater the larger is the initial
anisotropy of the plasma. As noted above, the linear stage ends when the system formed a pronounced
current structure consisting of a system of current beams (Fig.\:\ref{F:Jz2D}, time slice $t =
150\,[1/\omega_{\mathrm{p}e}]$), and when the particles become significantly magnetized in average
(mean Larmor radius is of the order of the current beam radius).

Further development of the WI is nonlinear and is accompanied by dynamic merging of equidirected
current filaments into larger-scale structures, that can be seen from a series of time slices shown
in Fig.\:\ref{F:Jz2D}. Along with that the magnitude of the magnetic field energy stabilizes. Here
we would note that this picture is quite similar to one considered in \cite{Lee.Lampe}, where the
authors study filament recombination which take place on the nonlinear stage of WI arising with
propagation of relativistic electron beam in background plasma (a theory of relativistic
coalescence has been developed in \cite{Medvedev, Polomarov}).

The merging continues until the velocity distribution of the electrons does not become an
equilibrium. In that context it is interesting to trace the evolution of the initial anisotropy of
the electron component during the development of WI, including at the non-linear stage. The
dependency of the anisotropy on the time is shown in Fig.\:\ref{F:Aniso}. From the figure one can
see that the process of growth and further stabilization of current density and magnetic field
energy density is accompanied by isotropization (decrease in the anisotropy) of the medium, caused
by collective processes of formation and restructuring of current filaments together with
competitive development of the unstable modes of the magnetic field.
\begin{center}
    \begin{minipage}[t]{\columnwidth}
        \refstepcounter{figure}
        \begin{center}
            \includegraphics[width=\columnwidth,keepaspectratio]
            {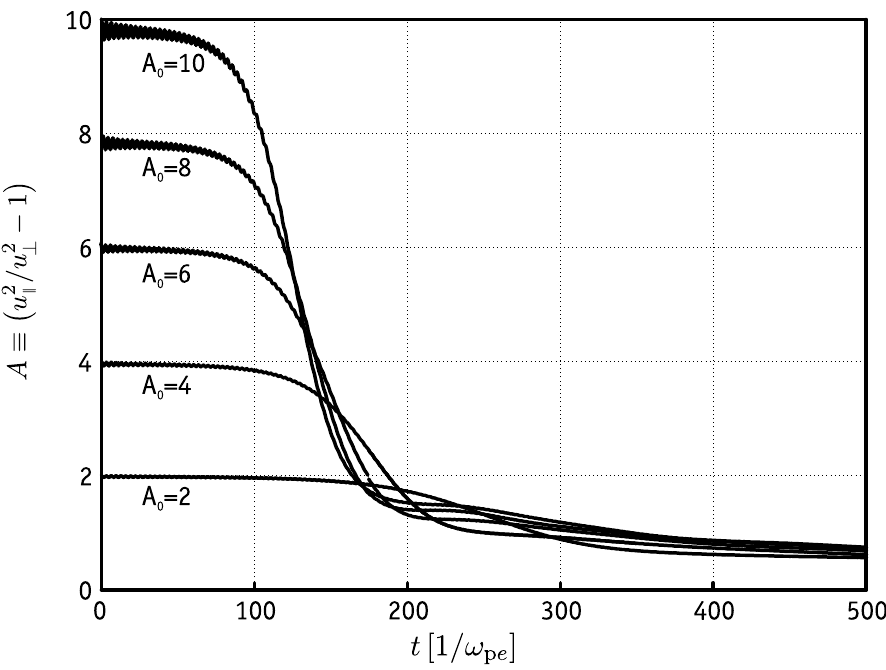}\label{F:Aniso}
        \end{center}
        \textit{Fig.\:\arabic{figure}.
        Evolution of the anisotropy for different initial $A_0$. The same setup as for
        Fig.\:\ref{F:WebelStepsLnBEnergy}.}
    \end{minipage}
\end{center}

It is interesting to note here an non-obvious fact that regardless of the initial value of the the
anisotropy, its value at the instability saturation stage tends to a nonzero threshold value (see
Fig.\:\ref{F:Aniso}).

The origin for that residual anisotropy is in the problem setup, and it is connected with an actual
finiteness of the perturbation spectrum of any system with periodic boundary conditions, due to the
fact that in such a system waves longer than linear size of the domain can not exist.

This conclusion is confirmed by the results of \cite{Lemons.Winske.Gary}, where it is shown that
for a limited system (in the sense stated above) the minimum level of the anisotropy at the
saturation stage of WI can be defined by the following expression:
\begin{equation*}\label{E:AMin}
    A_{\mathrm{min}} = (k_{\mathrm{min}}c/\omega_{\mathrm{p}e})^2.
\end{equation*}
In our case the minimal wave number allowed in the system is
$k_{\mathrm{min}}c/\omega_{\mathrm{p}e} = 2\pi/L_{x,y} = 2\pi/25 = 0.25$, which gives
$A_{\mathrm{min}} = 0.063$. Some discrepancy of the experimentally observed levels of residual
anisotropy (Fig. \ref{F:Aniso}) and its theoretically predicted value can be explained by the fact,
that at time $t = 500\,[1/\omega_{\mathrm{p}e}]$ the dominant mode of the instability has not yet
reached the limit determined by the period of the system. As stated above, one of the objectives of
these experiments was to clarify the dependency of the basic characteristics of the
instability---its growth rate, the characteristic time and the maximum energy density of the
magnetic field, of the initial anisotropy of plasma.

In this connection let's return to Fig.\:\ref{F:WebelStepsLnBEnergy}, from which one can see the
expected effect: the increase of the maximum field energy density and instability growth rate with
the increase of initial anisotropy. Less trivial is the profile of this dependency shown in
Fig.\:\ref{F:Max1D2DBEn}: fast increase of the maximum energy density on the interval $A_0 < 10$,
followed by an equally fast drop-off with $A_0 > 10$.

Moreover the value of $A_0$ around 25 can be considered as an upper threshold of instability in the
sense that further increase in the initial anisotropy of electron component has virtually no effect
on the basic characteristics of the WI---the maximum value of magnetic field energy density (in
case of fixed value of accentuated thermal velocity $u_z$). The observed effect and its
quantitative characteristics are confirmed by analytical expression which can be obtained on the
basis of the expression derived in the work \cite{Lemons.Winske.Gary}:
\begin{equation}\label{E:WBvsKin}
    \frac{w_B}{n_0 T_{z0}} = \frac{1}{3}\left(\frac{A}{A+1}\right)\left(\frac{1}{A+1} -
    \frac{1}{A_0+1}\right),
\end{equation}
where $T_z$ is the temperature of accentuated electron component along $z$-axis.

In the just mentioned work the expression (\ref{E:WBvsKin}) has been used as an estimation of
evolution of the magnetic field energy density, where the values of $A$ were being taken from the
numerical simulation.

However, in our context, it can be interpreted as an analytical dependency of the magnetic field
energy density (in units of $n_0 T_z$) of the current value of anisotropy $A$ with its initial
value given as $A_0$. Considering (\ref{E:WBvsKin}) that way, its easy to get an interesting for us
estimation of the maximum energy of WI (or, equivalently, the maximum proportion of the energy of
accentuated component which can be converted into field energy) and also the explanation of the
existence of upper threshold of WI on $A_0$ mentioned above. Indeed, differentiating the right hand
side of (\ref{E:WBvsKin}) with respect to $A$ and equating the derivative to zero we obtain:
\begin{equation*}\label{E:WBvsKinDiff}
    \frac{A_0 - A(2 + A_0)}{3(A+1)^2(A_0+1)} = 0.
\end{equation*}
Where from we get the value of the anisotropy coefficient which delivers the maximum to the
expression (\ref{E:WBvsKin}) with fixed value of $A_0$:
\begin{equation*}\label{E:WBvsKinAExtr}
    A_{\mathrm{extr}} = \frac{A_0}{2 + A_0}\,.
\end{equation*}
Substituting the resulting value of $A_{\mathrm{extr}}$ back into (\ref{E:WBvsKin}), we find the
desired dependence of the maximum magnetic field energy on the value of $A_0$:
\begin{equation}\label{E:WBmax}
    w_{B\mathrm{max}} = \frac{n_0 T_{z0} A_0^2}{12 (1 + A_0)^2} = \frac{w^\mathrm{kin}_{z0} A_0^2}{6 (1 +
    A_0)^2}\,,
\end{equation}
where $w^\mathrm{kin}_{z0}$ is the $z$-component of the average over domain initial kinetic energy
density of electrons. Dependency (\ref{E:WBmax}) together with the experimental values is shown in
Fig. \ref{F:Max1D2DBEn}.

\begin{center}
    \begin{minipage}[t]{\columnwidth}
        \refstepcounter{figure}
        \begin{center}
            \includegraphics[width=\columnwidth,keepaspectratio]
            {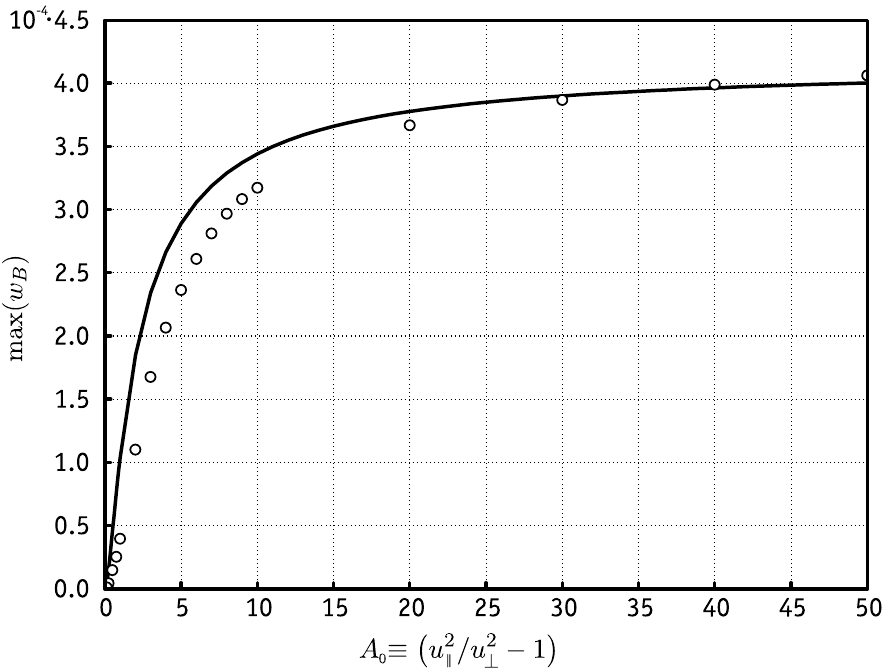}\label{F:Max1D2DBEn}
        \end{center}
        \textit{Fig.\:\arabic{figure}.
        The dependency of the maximum average magnetic field energy density of the initial anisotropy
        at constant accentuated thermal velocity $u_z = 0.1\,[c]$. The same setup as for
        Fig.\:\ref{F:WebelStepsLnBEnergy}.}
    \end{minipage}
\end{center}

Analysis of the last expression gives a number of interesting provisions which are in good
agreement with computer simulation experiments.

First, $w_{B\mathrm{max}}$ has a limit in case of fixed values of $n_0$, $T_{z0}$ ($u_z$) and
$A_0\rightarrow\infty$ (the last one actually means $u_x, u_y\rightarrow0$)
\begin{equation*}\label{E:WBmaxLim}
    w^{\lim}_{B\mathrm{max}} =  \frac{n_0 T_{z0}}{12} = \frac{w^\mathrm{kin}_{z0}}{6}\,,
\end{equation*}
corresponding to the maximum possible energy density of WI. Particularly, in our case, where $n_0
T_{z0} = 0.005\,[n_e m c^2]$), $w^{\lim}_{B\mathrm{max}}$ equals to $4.1\cdot10^{-4}\,[n_e m c^2]$,
which, as it can be seen in Fig. \ref{F:Max1D2DBEn}, is in good agreement with the numerical data.

Second, observed in experiments, the upper threshold of WI with respect to $A_0$ is determined by
the value of $A_0$, starting from which the fraction $A_0^2/(1 + A_0)^2$ becomes close to unity. It
is easy to see that the value of $A_0$ close to 25 meets this requirement (indeed, if $A_0 = 25$
the fraction equals to 0.925, and if $A_0 = 50$, i.e. increased by factor of two, it only slightly
increases up to 0.96).

Third, the portion of kinetic energy stored in the accentuated component of the system which goes
for development of Weibel instability depends on the degree of the initial anisotropy ($A_0$) and
it can not be higher than 1/6. For instance, in our experiments it varies from $12.4\%$ (for $A_0 =
9$) down to $4.4\%$ (for $A_0 = 2$). These data are consistent both with the analytical predictions
obtained from the expression (\ref{E:WBmax}) and with the upper limit of the converted anisotropy
energy (around 10\%) obtained in \cite{Morse.Nielson}, where 2D simulation has been performed in
the frame of full electromagnetic model of plasma.

If we define the characteristic time of the instability ($T_I$) as the time needed to reach the
maximum of magnetic field energy density, then from the same Fig. \ref{F:WebelStepsLnBEnergy} one
can easily estimate both the value of $T_I$ as well as it's dependence of the initial anisotropy
$A_0$.

Finally, we note good agreement between the numerical value of the dimensionless instability growth
rate (0.034) with the one predicted by the linear theory (0.037) for the setup considered above
where $A_0 = 9$ and $u_z=0.1\,[c]$.

\section*{Conclusion}
Thus, the performed computer experiments allowed to supplement the classical picture of the Weibel
instability. Particularly, we obtained the dependency of the characteristic time of its development
and the maximum magnetic field energy density of the initial value of the anisotropy parameter.
Also we have traced the dynamic restructuring of the current filaments accompanying nonlinear stage
of instability saturation, as well as the evolution of the initial anisotropy of the electron
component of the plasma.

As a further work it is planned to perform a numerical study of the dynamics and generation
mechanisms of two-species Weibel instability, which implies a kinetic representation not only of
the electron but also the ion plasma component on the time scale by orders of magnitude grater than
it is for a single-species WI.

Simulations have been performed on SKIF MSU "Chebyshev" cluster of Moscow State University Research
Computing Center.

\section*{Acknowledgments}
The authors consider it their pleasant duty to thank Acad. L.\,M.~Zelenyi and Dr. H.\,V.~Malova
from Russian Space Research Institute for numerous constructive discussions of the results
presented in this work.

\end{multicols}

\begin{center}
    \begin{minipage}[t]{\columnwidth}
        \refstepcounter{figure}
        \begin{center}
            \includegraphics[width=\columnwidth,keepaspectratio]
            {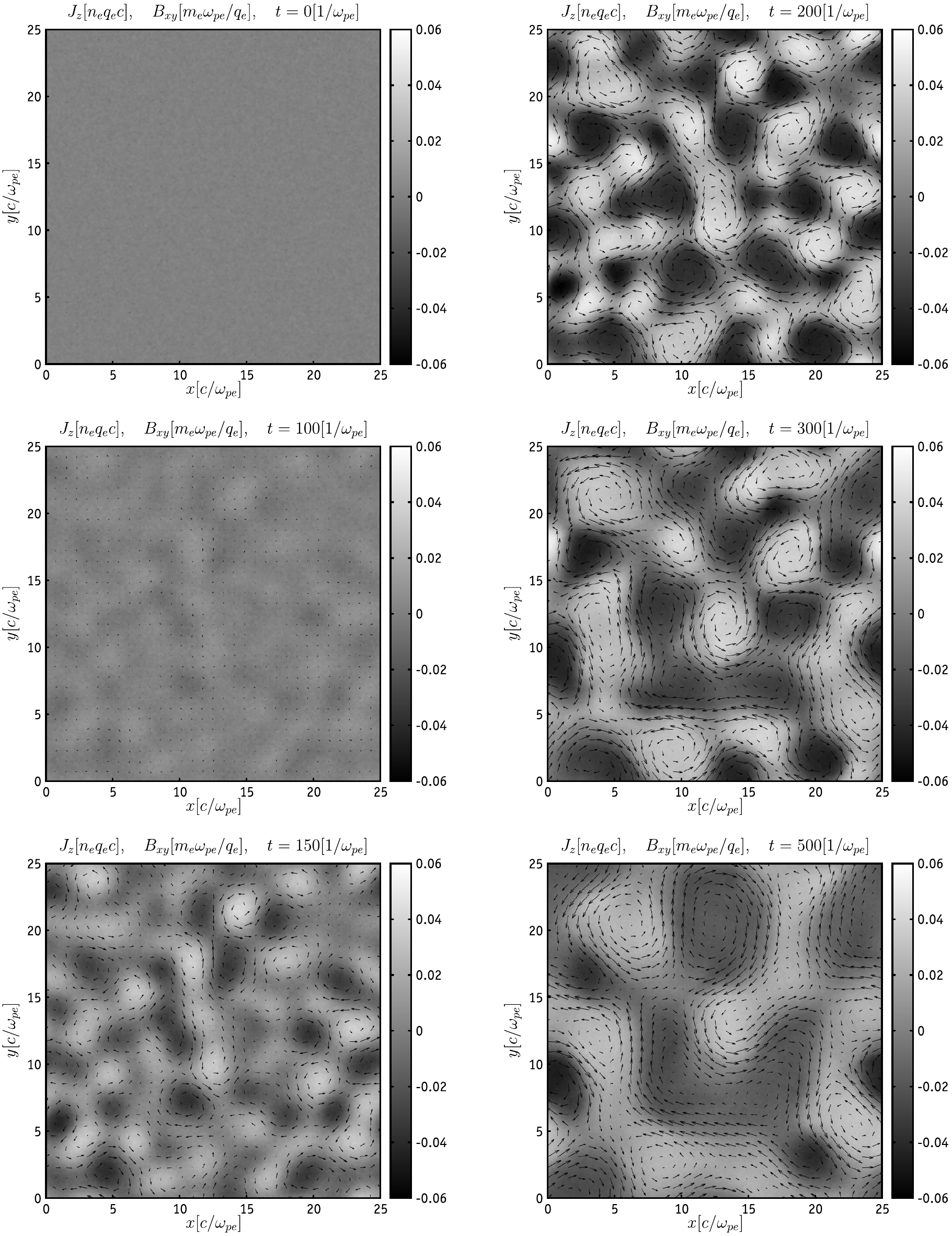}\label{F:Jz2D}
        \end{center}
        \textit{Fig.\:\arabic{figure}.
        The current density $J_z$ and the magnetic field $\vec{B}_{xy}$ at different time moments
        for 2.5-dimensional setup of WI in case $A_0 = 9$. Computational domain
        $L_x = L_y = 25\,[c/\omega_{\mathrm{p}e}]$; $u_z = 0.1\,[c]$;
        mesh size $256\times256$; 1000 particles of each specie per cell;
        time step $\tau = 0.25\,[1/\omega_{\mathrm{p}e}].$}
    \end{minipage}
\end{center}

\begin{thebibliography}{11}
    \parskip 0pt
    \parsep 0pt
    \itemsep 3pt
    \bibitem{Weibel.article}
        \textit{Weibel E.\,S.} Spontaneously Growing Transverse Waves in a Plasma Due to an Anisotropic Velocity
        Distribution. // Phys. Rev. Lett., v. \textbf{2}, 1959, 83--84.
    \bibitem{Pukhov}
        \textit{Pukhov A., Meyer-ter-Vehn J.} Relativistic Magnetic Self-Channeling of Light in Near-Critical
        Plasma: Three-Dimensional Particle-in-Cell Simulation. // Phys. Rev. Lett.,
        v. \textbf{76}, №21, 1996, 3975--3978.
    \bibitem{Yoon.ionWI}
        \textit{Yoon P.\,H., Lui A. T. Y.} Nonlocal ion-Weibel instability in the geomagnetic tail.
        // J.: Geophys. Res., v. \textbf{101}, №.A3, 1996, 4899--4906.
    \bibitem{Davidson.MWI}
        \textit{Davidson R.\,C., Startsev E.\,A., Kaganovich I., Qin H.} Multispecies Weibel Instability
        for Intense Ion Beam Propagation Through Background Plasma. // PAC 2005. Proceedings,
        2005, 1952--1954.
    \bibitem{Tsintsadze.Shukla}
        \textit{Tsintsadze L.\,N., Shukla P.\,K.} Weibel instabilities in dense quantum plasmas.
        J. Plasma Phys. v. \textbf{74}, No. 4, 2008, 431--436.
    \bibitem{Borodachev.model}
        \textit{Borodachev L.\,V., Mingalev I.\,V., Mingalev O.\,V.} Vlasov--Darwin System.
        // Encyclopedia of Low Temperature Plasma (Series B), v. \textbf{VII}. M.:~"Janus-K", 2008, 136--146.
    \bibitem{Mikhaylovskiy.plasma}
        \textit{Mikhailovsky A.\,B.} Theory of Plasma Instabilities. M.:~Atom Press, 1975.
    \bibitem{Morse.Nielson}
        \textit{Morse R.\,L., Nielson C.\,W.} Numerical Simulation of the Weibel Instability in One
        and Two Dimensions. // The Physics of Fluids, v. \textbf{14}, No. 11, 1971, 830--840.
    \bibitem{Frank}
        \textit{Frank-Kamenetskii D.\,A.} Lectures on plasma physics. M.:~Atom Press, 1964.
    \bibitem{Kalitkin} \textit{Kalitkin N.\,N.} Numerical Methods. M.:~"Nauka", 1978.
    \bibitem{Darwin}
        \textit{Darwin C.\,G.} Dynamical Motions of Charged Particles, \textit{Phil. Mag.}, \textbf{39}, 1920,
        537-546.
    \bibitem{Nielson}
        \textit{Nielson C.\,W., Lewis H.\,R.} Particle-Code Methods in the Nonradiative Limit.
        \textit{Meth. Comput. Phys.}, v. \textbf{16}, 367--388, 1976.
    \bibitem{Hockney.book}
        \textit{Hockney R.\,W., Eastwood J.\,W.} Computer simulation using particles. CRC Press,
        1988.
    \bibitem{Lee.Lampe}
        \textit{Lee R., Lampe M.} Electromagnetic Instabilities, Filamentation, and Focusing
        of Relativistic Electron Beams. // Phys. Rev. Lett., v. \textbf{31}, Issue 23, 1973,
        1390--1393.
    \bibitem{Medvedev}
        \textit{Medvedev M.\,V., Fiore M., Fonseca R.\,A., Silva L.\,O., Mori W.\,B.}
        Long-Time Evolution of Magnetic Fields in Relativistic Gamma-Ray Burst Shocks.
        // The Astrophysical Journal, v. \textbf{618}, 2005, L75--L78.
    \bibitem{Polomarov}
        \textit{Polomarov O., Kaganovich I., Shvets G.}
        Merging of Super-Alfv\'{e}nic Current Filaments during Collisionless Weibel Instability
        of Relativistic Electron Beams. // Phys. Rev. Lett., v. \textbf{101}, 2008, 175001.
    \bibitem{Lemons.Winske.Gary}
        \textit{Lemons D.\,S., Winske D., Gary S.P.} Nonlinear theory of the Weibel instability.
        // J. Plasma Phys., v. \textbf{21}, part 2, 1979, 287--300.
\end{thebibliography}
\end {document}